\documentclass[twoside]{article}
\usepackage{fleqn,espcrc2}

\usepackage{graphicx}
\def\kpnn{K^+\rightarrow\pi^+\nu\bar\nu}

\def\klpn{K_{\rm L}\rightarrow\pi^0\nu\bar\nu}

\def\klee{K_{\rm L}\rightarrow\pi^0 e^+e^-}
\newcommand{\eps}{\varepsilon}
\newcommand{\epe}{\varepsilon'/\varepsilon}
\renewcommand{\Re}{\mathrm{Re}}
\renewcommand{\Im}{\mathrm{Im}}
\newcommand{\nn}{\nonumber}

\title{
\vspace{-1cm} \rightline{\normalsize{TUM-HEP-393/00}}
\rightline{\normalsize{ROME1-1301/00}}
\vspace{.5cm}
$\epe$ and Rare $K$ Decays in the Standard Model and Supersymmetry}

\author{Luca Silvestrini\address{Dip. di Fisica, Univ. di
    Roma ``La Sapienza'' and INFN, sez. di Roma, P.le A. Moro, I-00185
    Roma, Italy}\thanks{Work supported in part by the MURST and by the
    German Bundesministerium f\"ur Bildung und Forschung under
    contracts 06 TM 874 and 05HT9WOA0.}\thanks{Talk given at the 4th
    International Conference On Hyperons, Charm And Beauty Hadrons,
    27-30 June 2000, Valencia, Spain.}} 
       
\begin{document}
\pagestyle{empty}

\begin{abstract}
  After briefly reviewing the status of $\epe$ in the Standard Model,
  I discuss SUSY contributions to $\epe$, $\kpnn$, $\klpn$ and
  $\klee$. While in the simplest case of the MSSM with Minimal Flavour
  Violation the main effect is a suppression of these transitions with
  respect to the Standard Model, large enhancements are possible in
  more general SUSY models, with interesting correlations among the
  different processes.
\end{abstract}

\maketitle

\section{Introduction}

After the new measurements of $\epe$ provided by the KTeV and NA48
collaborations, 
\begin{eqnarray}
  \Re\, \epe &=& \left( 28.0 \pm 4.1 \right) \cite{KTeV}\,,\nn \\
  \Re\, \epe &=& \left( 14.0 \pm 4.3 \right) \cite{NA48}\,,
  \label{eq:epexp}
\end{eqnarray}
some very interesting theoretical questions have arisen: i) Does the
Standard Model (SM) prediction for $\epe$ agree with the above
experimental results? ii) What kinds of new physics can improve the
agreement with the experiment? iii) In particular, in the framework of
SUSY models, in what region of SUSY parameter space can large
contributions to $\epe$ arise? iv) What is the impact of
$(\epe)_\mathrm{exp}$ on predictions for rare $K$ decays, both in the
SM and beyond?

In this talk, I will briefly discuss the above questions.
Unfortunately, most of them depend very strongly on poorly understood
hadronic dynamics. Therefore, for most of the above questions no
conclusive answers can be provided at present.

\section{A quick look at $\epe$ in the SM}
\label{sec:SM}

The basic formula for $\epe$ is given by
\begin{equation}
\frac{\varepsilon^\prime}{\varepsilon} = 
\Im \lambda_t\cdot F_{\varepsilon'},
\label{eq:epe1}
\end{equation}
where
\begin{equation}
F_{\varepsilon'} = 
\left[ P^{(1/2)} - P^{(3/2)} \right],
\label{eq:epe2}
\end{equation}
with
\begin{eqnarray}
P^{(1/2)} & = & r \sum y_i \langle Q_i\rangle_0
(1-\Omega_{\eta+\eta'})~,
\label{eq:P12} \\
P^{(3/2)} & = &\frac{r}{\omega}
\sum y_i \langle Q_i\rangle_2~,
\label{eq:P32}
\end{eqnarray}
and we have neglected the small phase difference between $\eps$ and
$\varepsilon^\prime$. Here
\begin{eqnarray}
&& \langle Q_i\rangle_I \equiv \langle (\pi\pi)_I | Q_i | K \rangle\,,
\nn \\
&& r = \frac{G_{\rm F} \omega}{2 |\eps| \Re A_0}~~~\mathrm{and}~~~ 
\omega = \frac{\Re A_2}{\Re A_0} \sim \frac{1}{22}  
\label{eq:repe}
\end{eqnarray}
(see however ref.~\cite{GV2} for possible isospin breaking effects in
$\omega$).
The complete listing of the four-fermion operators $Q_i$ is given for
example in \cite{EP99}. The Wilson coefficient functions $y_i(\mu)$
were calculated including the complete next-to-leading order (NLO)
corrections in \cite{NLO}. The details of these
calculations can be found there and in the review \cite{BBL}.

It is customary (and convenient) to write $\langle Q_i \rangle_I \equiv
(\langle Q_i\rangle_I)_\mathrm{VIA} B_i^{(I)}$, where the $B_i^{(I)}$
parameterize the deviation of the full matrix elements from the Vacuum
Insertion Approximation (VIA). However, by doing so one introduces a
strong dependence on the mass of the strange quark $m_s$, which could be
avoided if one were able to compute directly the full matrix
element \cite{romams}. 

The sum in (\ref{eq:P12}) and (\ref{eq:P32}) runs over all
contributing operators. $P^{(3/2)}$ is fully dominated by electroweak
penguin contributions. $P^{(1/2)}$ on the other hand is governed by
QCD penguin contributions. Isospin breaking in the quark masses ($m_u
\not= m_d$) induces a further $\Delta I=3/2$ contribution. The latter
effect is described by $\Omega_{\eta+\eta'}$.

The dominant contributions in the SM come from the QCD-penguin
operator $Q_6$ and from the electroweak-penguin operator $Q_8$, as
discussed in refs. \cite{eppRoma,eppMonaco}, where further details as well
as the full expressions for $\epe$ can be found. $\epe$ can then be
written as
\begin{displaymath}
  \epe=\Im \lambda_t F_\varepsilon^\prime
  \left(m_t,\alpha_s,m_s,B_6^{(1/2)},B_8^{(3/2)},\Omega_{\eta+\eta'}\right),
\end{displaymath}
where only the main dependencies have been shown. $\Im \lambda_t
\equiv \Im V_{td}V_{ts}^*$ has to be determined using the information
on the Unitarity Triangle (UT) coming from $V_{ub}$, $V_{cb}$,
$\varepsilon_K$ and $\Delta M_{B_{d,s}}$. Since the $\Delta I=1/2$
contribution of $Q_6$ and the $\Delta I=3/2$ contribution of $Q_8$
have opposite signs and cancel each other to a large extent, the final
result for $\epe$ depends very strongly on $B_6^{(1/2)}$ and
$B_8^{(3/2)}$. Unfortunately, while $B_8^{(3/2)}$ can be reliably
computed with lattice QCD (see ref. \cite{B8} for an up-to-date
review), at present there is no computation of $B_6^{(1/2)}$ from
first principles. Although some suggestions on how to compute
$B_6^{(1/2)}$ on the lattice, including Final State Interactions
(FSI), have been recently made \cite{latnonlep}, we have to face the
fact that no reliable prediction for $B_6^{(1/2)}$ is currently
available. This precludes to a large extent the possibility of a
precise computation of $\epe$ in the SM. If the value of $B_6^{(1/2)}$
is close to unity, then the SM prediction is smaller than the
experimental value of $\epe$. For instance, using $\hat B_K=(0.80 \pm
0.15)$, $m_s(m_c)=(130 \pm 25)$ MeV, $B_6^{(1/2)}=(1.0 \pm 0.3)$,
$B_8^{(3/2)}=(0.8 \pm 0.2)$ and $\Omega_{\eta+\eta'} = (0.16\pm0.03)$
as recently obtained in ref. \cite{ECKER99} (see however \cite{GV} for
an estimate of the uncertainties in isospin breaking effects),
one gets the following prediction for $\epe$ \cite{EP99,EPPMSSM}:
\begin{eqnarray}
  (\epe)_\mathrm{NDR}&=&(9.2^{+6.8(19.3)}_{-4.0(6.2)})\times 10^{-4}\,,
  \nn \\
  (\epe)_\mathrm{HV}&=&(6.4^{+5.3(14.8)}_{-3.1(4.9)})\times 10^{-4}\,.
  \label{eq:eppnoi}
\end{eqnarray}
Since $B_6^{(1/2)}$ is certainly the less known ingredient in
(\ref{eq:eppnoi}), it is instructive to compute $\epe$ leaving
$B_6^{(1/2)}$ as a free parameter. One gets, with the same choice of
parameters as above,
\begin{equation}
  \label{eq:eppB6}
  (\epe)_\mathrm{NDR}=(16.2^{+8.9}_{-5.1}B_6^{(1/2)}
  -7.8^{+2.0}_{-2.2}) \times 10^{-4}.
\end{equation}

\begin{table}[t]
\begin{center}
\caption[]{ Results for $\epe$ in the SM in  units of $10^{-4}$.
\label{tab:epeSM}}
\begin{tabular}{||c|c||}\hline \hline
  {\bf Reference}&  $\epe~[10^{-4}]$ \\ \hline
Munich
\cite{EP99}&  $9.2^{+6.8}_{-4.0}$ (MC) \\
Munich
\cite{EP99}&   $1.4\to 32.7$ (S) \\
\hline
Rome
\cite{CM00,ROMA99}& $8.1^{+10.3}_{-9.5}$ (MC) \\
Rome
\cite{CM00,ROMA99}& $-13.0\to 37.0$ (S) \\
\hline
Trieste
\cite{BERT98}&  $22\pm 8 $ (MC) \\
Trieste
\cite{BERT98}&  $9\to 48 $ (S) \\
\hline
Dortmund
\cite{Dortmund}&    $6.8\to 63.9$ (S) \\
\hline
Montpellier
\cite{Narison}&  $ 24.2\pm 8.0$ \\
\hline
Granada-Lund
\cite{Prades}&   $ 34\pm 18$ \\
\hline
Dubna-DESY
\cite{BEL} 
&  $-3.2 \to 3.3$ (S) \\
\hline
Taipei
\cite{Tajpei}&  $7 \to 16$ \\
\hline
Barcelona-Valencia 
\cite{PAPI99}&  $17 \pm 6$ \\
\hline \hline
\end{tabular}
\end{center}
\end{table}

Other recent theoretical estimates of $\epe$ are reported in table
\ref{tab:epeSM}. They differ in the evaluation of the relevant
hadronic matrix elements. The Rome group is using lattice results
wherever they are available \cite{CM00,ROMA99}. Their results are in
good agreement with the estimates in eq. (\ref{eq:eppnoi}). The
Dortmund group is computing higher order terms in the chiral and $1/N$
expansions, which unfortunately contain quadratic divergences. This
makes it very difficult to reliably match their estimates of the
hadronic matrix elements with the short distance computation
\cite{Dortmund}. The difficulty in the matching with the short
distance part is also a limitation for Chiral Quark Model computations
such as the one performed by the Trieste group \cite{BERT98}. The same
problems arise in the Extended Nambu-Jona-Lasinio Model used by the
Dubna group \cite{BEL}. The Taipei group has applied generalized
factorization to the computation of $K \to \pi \pi$ amplitudes
\cite{Tajpei}. Being just a general parameterization of nonleptonic
decay amplitudes, generalized factorization provides no dynamical
information. In order to obtain predictions for $\epe$, some
simplifying assumptions have to be made in order to reduce the number
of unknown parameters. Unfortunately, these assumptions cannot be
justified from first principles, nor can they be verified
experimentally. QCD sum rules have been used in the Montpellier
estimate of $\epe$ \cite{Narison}. The Granada-Lund group has used the
X-boson method to match the short-distance computation to an estimate
of the matrix elements performed at the NLO in the $1/N$ expansion in
the chiral limit \cite{Prades}.

What can we learn from the results in table \ref{tab:epeSM}?  First of
all, one should stress that, although quite improbable, it is not
impossible to obtain $\epe \sim 2 \times 10^{-3}$ using the Munich and
Rome analyses. However, clearly a larger value of $B_6^{(1/2)}$ would
improve the agreement with the experimental data. It is tempting to
suppose that there is a common origin for the large value of $\epe$
and the $\Delta I=1/2$ rule. However, to relate these two quantities
one has to make some dynamical assumption \cite{charming} or to use
some model, such as the ones discussed above. It is quite interesting
that most of the model-dependent computations presently available, in
which a connection between $\epe$ and the $\Delta I=1/2$ rule is
established, find central values for $\epe$ in the ballpark of the
experimental value. However, more theoretical progress is needed to
put this connection on solid grounds. Another interesting
possibility, recently proposed in ref. \cite{PAPI99}, is that the
large FSI in the $\Delta I=1/2$ channel enhances the $\Delta I=1/2$
matrix elements and effectively gives $B_6^{(1/2)} \sim
1.5$. Unfortunately, a reliable quantitative evaluation of FSI effects
in $\langle Q_6\rangle_{(1/2)}$ is not available yet, since the
method applied in ref. \cite{PAPI99} to resum the pion bubble diagrams
suffers from systematic uncertainties and only applies to the lowest
order in the $1/N$ expansion (see refs. \cite{noicritical,PAPI00} for
further details and different opinions on this subject).  

\section{$\epe$ and Rare $K$ Decays in SUSY with Minimal Flavour
  Violation}
\label{sec:MFV}

From the above discussion, it is clear that there is certainly still
room for a large New Physics contribution to $\epe$. It is therefore
worthwhile to explore possible SUSY effects in $\epe$.  Let us start
by considering a very constrained class of SUSY models, the MSSM with
Minimal Flavour Violation (MFV), where no new phase and no new source
of Flavour Violation is introduced. In this kind of extensions of the
SM, all flavour and CP violation is governed by the CKM matrix.  Rare
$K$ decays and $\epe$ can then be affected by SUSY in two ways.  First
of all, new contributions to the $\Delta S=2$ amplitude, which in this
class of models turn out to be always positive, modify the UT fit,
resulting in smaller values for $\Im \lambda_t$ and $\Re \lambda_t$,
and therefore suppressing $\epe$ and rare $K$ decays.  Then, there are
new contributions to the $\Delta S=1$ amplitudes. In a few cases,
these can enhance the SM amplitudes and overcompensate the suppression
due to $\Im \lambda_t$ and $\Re \lambda_t$, resulting in larger values
of $\epe$ and larger BR's for rare $K$ decays. In most cases, however,
the net effect is a suppression with respect to the SM. I summarize
here the results of a recent detailed analysis in this class of models
\cite{EPPMSSM}, taking into account all the available constraints
(direct searches of SUSY particles, BR$(b \to s \gamma)$, precision
electroweak data, lower bounds on the neutral Higgs mass):
\begin{itemize}
\item In most of the allowed parameter space, the overall effect is a
  suppression of $\epe$, mainly due to the positive contributions to
  $\eps$, as already found in ref. \cite{GG};
\item The strongest suppression of $\epe$, of about a factor of two,
  is obtained for low $\tan \beta$, low $m_{H^\pm}$, for a light,
  mainly right-handed stop and for large splitting between the stops;
\item A modest ($7 \%$) enhancement of $\epe$ can be obtained for
  relatively large $\tan \beta$ and $m_{H^\pm}$ and for a light
  chargino;
\item The same pattern emerges for rare $K$ decays:
  $\mathrm{BR}(\kpnn)$, $\mathrm{BR}(\klpn)$ and $\mathrm{BR}(\klee)$
  are mainly suppressed with respect to the SM prediction, up to
  approximately a factor of two, while only very modest enhancements
  are possible.\footnote{Here and in the following, I indicate by
    $\mathrm{BR}(\klee)$ the direct CP violating contribution to this
    decay.}
\end{itemize}
The minimal and maximal values of the ratios $T$, defined as the
prediction in the MSSM with MFV normalized to the SM central value,
are reported in table \ref{tab:MFV} (see ref. \cite{EPPMSSM} for
details).

\begin{table}[t]
\caption[]{The minimal and maximal values of the the ratios $T$
without constraints and with all constraints taken into account.
\label{tab:MFV}}
\begin{center}
\begin{tabular}{|c|c|c|c|c|}
\hline
& \multicolumn{2}{c|}{{\rm No Constr.}} &
  \multicolumn{2}{c|}{{\rm All Constr.}} \\
\hline
$T $ & {\bf min} & {\bf max} & {\bf min} & {\bf max} \\
\hline
$T(\Im \lambda_t)$ & $0.57$ & $1.00$ & $0.66$ & $1.00$    \\
$T(\Re \lambda_t)$ & $0.78$ & $1.00$ & $0.81$ & $1.00$     \\
$T(\epe)$ & $0.42$ & $1.07$ & $0.53$ & $1.07$ \\
$T(\kpnn)$ & $0.59$ & $1.09$ & $0.65$ & $1.02$  \\
$T(\klpn)$ & $0.28$ & $1.12$ & $0.41$ & $1.03$     \\
$T(\klee)$ & $0.33$ & $1.10$ & $0.48$ & $1.10$  \\
\hline
\end{tabular}
\end{center}
\end{table}

\section{$\epe$ and Rare $K$ Decays in General SUSY Models}
\label{sec:General}

The situation drastically changes when one abandons the assumption of
MFV and allows for the most general flavour and CP structure in the
soft SUSY breaking terms. In this case, there are in general new
independent contributions to $\Delta S=2$, $\Delta B=2$ and $\Delta
S=1$ transitions, with arbitrary phases. Furthermore, the so-called
magnetic moment and chromomagnetic operators,
negligible in the SM, can give large contributions to $\epe$ and
$\klee$. The new flavour and CP violating parameters introduced in
these models are subject to quite stringent constraints coming from
the available FCNC data \cite{GMS,GGMS}. However, it is interesting
that, compatibly with these tight constraints, large contributions to
$\epe$ and rare $K$ decays are still possible \cite{GGMS,CI}. A
combined analysis shows that interesting correlations can be
established between these new contributions \cite{noirare}. 
In the rest of this talk, I will briefly review three possible
scenarios in which large SUSY contributions to $\epe$ and rare $K$
decays are present.

\subsection{Scenario I: Enhanced $Z^0$ Penguins}
\label{sec:z0}

Colangelo and Isidori pointed out some time ago that a large $\bar s d
Z$ vertex can be induced in the presence of large $\tilde t_R$-$\tilde
s_L$ and $\tilde t_R$-$\tilde d_L$ mixings \cite{CI}. The $\bar s d Z$
vertex has a strong impact not only on the decays $\kpnn$, $\klpn$ and
$\klee$, but also on $\epe$ \cite{BS99}.  The experimental results on
$\epe$ then constrain the possible enhancement of rare decays. The
effective $\bar s d Z$ vertex can be written as
 \begin{displaymath}
%  \label{eq:Wds}
  {\cal H}^{sdZ}_{\rm eff} = -\frac{G_F}{\sqrt{2}} \frac{e}{ \pi^2} M_Z^2
  \frac{\cos \Theta_W}{\sin \Theta_W} Z_{ds} \bar s_L \gamma_\mu
   Z^\mu d_L \,+\, {\rm h.c.}, 
\end{displaymath}
where 
\begin{equation}
  \label{eq:Wsm}
  Z_{ds} = \lambda_t C_0(x_t)+\Lambda_t~.
\end{equation}
Here the first term on the r.h.s is the Standard Model contribution
(evaluated in the 't Hooft-Feynman gauge)
and $\Lambda_t$ is an effective SUSY coupling, whose definition is
given in ref. \cite{noirare}, where further details can be found.
Since $Z_{ds}$ gives a negative contribution to $\epe$, a large value of
$\epe$ favours negative values of $Z_{ds}$ (i.e. with the opposite
sign compared to the SM). In table \ref{tab:rare1}, we report the
upper bounds on rare decays for $\epe=2 \times 10^{-3}$
\cite{noirare}. The upper bounds depend on the sign  of
$\Lambda_t$ and on the value of the SM coupling $\lambda_t$.

\begin{table}[t]
\begin{center}
\caption{Upper bounds for the branching ratios 
   of the rare decays $\klpn$, $\klee$ and
   $\kpnn$.
   The results have been  obtained 
   for $\lambda_t$ in the SM range (column I) and for $\lambda_t$ in
   the unitarity range (column II) by imposing 
   $\Re(\epe)\ge 2  \cdot 10^{-3}$ and $\Re(\epe)\le 2  \cdot 10^{-3}$
   for $\Im \Lambda_t>0$ and $\Im \Lambda_t<0$ respectively,  
   with $B_8^{(3/2)}=0.6 (1.0)$. The last column contains the upper
   bounds in the SM.
  \label{tab:rare1}}
  \begin{tabular}{|l|c|c|c|}
  \hline
  $\Im \Lambda_t>0$
  & I & II  & SM  \\ \hline
  $10^{10}~ {\rm BR}(\pi^0 \nu \bar \nu)$ & $0.7~(0.4)$ & $0.9~(0.5)$
  & 0.4 \\   
  $10^{11}~{\rm BR}(\pi^0 e^+ e^-)$  & $1.1~(0.7)$ & 
        $1.3~(0.8)$ & 0.7  \\ 
   $10^{10}~{\rm BR}(\pi^+ \nu \bar \nu)$   & $1.7~(1.7)$ &
  $2.0~(1.9)$ & 1.1  \\  
 \hline 
   $\Im \Lambda_t<0$ &
   I & II  & SM  \\ \hline 
   $10^{10}~ {\rm BR}(\pi^0 \nu \bar \nu)$ & $0.8~(0.4)$ & 
    $4.0~(3.8)$ & 0.4 \\  
   $10^{11}~ {\rm BR}(\pi^0 e^+ e^-)$  & $2.0~(0.7)$ & 
    $5.9~(5.7)$ & 0.7  \\  
   $10^{10}~ {\rm BR}(\pi^+ \nu \bar \nu)$ & $1.7~(1.7)$ & 
    $2.7~(2.6)$ & 1.1  \\
 \hline 
  \end{tabular}
\end{center}
\end{table}

\subsection{Scenario II: Enhanced Chromomagnetic Penguin}
\label{sec:chromo}

Another interesting possibility arises in models in which there is a
sizable $\tilde s_R$-$\tilde d_L$ mixing. In this kind of models, a
large contribution is induced to the magnetic and chromomagnetic
operators
\begin{eqnarray*}
Q^\pm_\gamma&=&\frac{Q_d e}{16 \pi^2}
        \left( {\bar s}_L \sigma^{\mu \nu} F_{\mu\nu} d_R \pm 
               {\bar s}_R \sigma^{\mu \nu} F_{\mu\nu} d_L \right), \\
Q^\pm_g&=&\frac{g}{16 \pi^2}
        \left( {\bar s}_L \sigma^{\mu \nu} t^a G^a_{\mu\nu} d_R \pm
               {\bar s}_R \sigma^{\mu \nu} t^a G^a_{\mu\nu} d_L \right).
\end{eqnarray*}
The chromomagnetic contribution to $\epe$ is unfortunately affected by
a large uncertainty in the evaluation of the hadronic matrix element.
However, it is interesting to notice that in SUSY models with
non-abelian flavour symmetries a chromomagnetic contribution to $\epe$
in the ball-park of $2 \times 10^{-3}$ can naturally arise \cite{MM}.
Large contributions of this kind can also arise in Left-Right SUSY
models \cite{mohapatra}. In some models, a large contribution to
$Q^-_g$ corresponds to a large contribution to $Q^+_\gamma$. In this
case, BR$(\klee)$ can be sizably enhanced over the SM prediction, up
to one order of magnitude (see ref. \cite{noirare} for a detailed
analysis).

\subsection{Scenario III: Isospin Breaking in Squark Masses}
\label{sec:isospin}

Finally, let me briefly mention an interesting possibility proposed
by Kagan and Neubert \cite{KN}. In SUSY models in which there is a
sizable $\tilde d_L$-$\tilde s_L$ mixing, and in which isospin is
broken in the squark masses $(m_{\tilde u} \ne m_{\tilde d})$, large
contributions to $\epe$ arise. The reason is that, due to the isospin
breaking in squark masses, SUSY-QCD box diagrams give rise to $\Delta
I=3/2$ transitions, which, as discussed in Sect. \ref{sec:SM}, enter
in $\epe$ with the enhancement factor $1/\omega \sim 22$. The $\tilde
d_L$-$\tilde s_L$ mixing is needed in order to generate $\Delta S=1$
transitions with this kind of diagrams.

\section{Conclusions}
\label{sec:concl}

Comparing the recent measurements of $\epe$ in eq.~(\ref{eq:epexp})
with the theoretical estimate of the Munich group in
eq.~(\ref{eq:eppnoi}), it is clear that, for the range of parameters
used in the Munich analysis in ref.~\cite{EPPMSSM}, the SM prediction
is somewhat lower than the experimental average. Within the framework
of the SM, many suggestions have been recently made that bring the
theoretical predictions closer to the experimental value. As I
discussed in Sect. \ref{sec:SM}, most of model-dependent estimates of
$\epe$ suggest a connection between the $\Delta I=1/2$ rule and the
large value of $\epe$. However, more theoretical work is needed to put
this intriguing possibility on solid grounds. On the other hand, a lot
of room is still available for large SUSY contributions to $\epe$. The
possibility of SUSY effects in $\epe$ is certainly very exciting, in a
world in which all other FCNC observables perfectly agree with SM
predictions. If one considers the ``simplest'' SUSY extension of the
SM, the MSSM with MFV, the results are however not so encouraging: for
both $\epe$ and rare $K$ decays, in most of the allowed parameter
space the net effect is a depletion of all these observables with
respect to the SM prediction. More interesting possibilities arise in
those SUSY extensions of the SM in which new sources of flavour and CP
violation are introduced. Remarkably, in most of these models the
potentially large contributions to $\epe$ are accompanied by even
larger contributions to rare $K$ decays, which can for example enhance
BR$(\klpn)$ and BR$(\klee)$ one order of magnitude above the SM upper
bound.  We therefore eagerly await for future progress both on the
theoretical side, with improved estimates of hadronic matrix elements,
and on the experimental side, with improved measurements of $\epe$ and
more stringent upper bounds and measurements of rare $K$ decays.

I am much indebted to A.J. Buras, G. Colangelo, P. Gambino, M.
Gorbahn, G. Isidori, S. J{\"a}ger and A. Romanino for a most enjoyable
collaboration. It is a pleasure to thank the organizers for the very
nice atmosphere, and in particular V. Gimenez for his warm
hospitality. I also thank A. Pich for discussions, I. Scimemi for
showing to me the best of Valencia and M. Ciuchini and E. Franco for
valuable comments and suggestions.

\end{document}